\documentclass[conference]{IEEEtran}
\IEEEoverridecommandlockouts
\usepackage{cite}
\usepackage{amsmath,amssymb,amsfonts}
\usepackage{algorithmic}
\usepackage{graphicx}
\usepackage{textcomp}
\usepackage{xcolor}
\graphicspath{{images/}}
\def\BibTeX{{\rm B\kern-.05em{\sc i\kern-.025em b}\kern-.08em
    T\kern-.1667em\lower.7ex\hbox{E}\kern-.125emX}}

\begin{document}

\title{Phishing in an Academic Community:\\
	A Study of User Susceptibility and Behavior\\
	\thanks{*Cyber Defense Lab}
}

\author{\IEEEauthorblockN{Alejandra Diaz, Alan T. Sherman,* Anupam Joshi}
	\IEEEauthorblockA{\textit{Department of Computer Science and Electrical Engineering} \\
		\textit{University of Maryland, Baltimore County (UMBC)}\\
		Baltimore Maryland, USA 21250\\
		\{adiaz1, sherman, joshi\}@umbc.edu}
}

\maketitle

\begin{abstract}
We present an observational study on the relationship between demographic factors and phishing susceptibility at the University of Maryland, Baltimore County (UMBC). 
In spring 2018, we delivered phishing attacks to 450 randomly-selected students on three different days (1,350 students total) to examine user click rates and demographics among UMBC's undergraduates. 
Participants were initially unaware of the study. 
We deployed the Billing Problem, Contest Winner, and Expiration Date phishing tactics. 
Experiment~1 claimed to bill students; Experiment~2 enticed users with monetary rewards; and 
Experiment 3 threatened users with account cancellation. 

We found correlations resulting in lowered susceptibility based on college affiliation, academic year progression, cyber training, involvement in cyber clubs or cyber scholarship programs, time spent on the computer, and age demographics. 
We found no significant correlation between gender and susceptibility. 
Contrary to our expectations, we observed greater user susceptibility with greater phishing knowledge and awareness. 
Students who identified themselves as understanding the definition of phishing had a higher susceptibility than did their peers who were merely aware of phishing attacks, with both groups having a higher susceptibility than those with no knowledge of phishing. 
Approximately 59\% of subjects who opened the phishing email clicked on its phishing link, and approximately
70\% of those subjects who additionally answered a demographic survey clicked.

\end{abstract}

\bigskip
\begin{IEEEkeywords}
Phishing, social engineering, cyber demographics, user susceptibility, cybersecurity,  \textit{Billing Problem} tactic, \textit{Contest Winner} tactic, \textit{Expiration Date} tactic.
\end{IEEEkeywords}

\section{Introduction}

Typically, the most important and devastating vulnerability a company can have is its very own people \cite{b6}. The human factor, or error, is responsible for 95\% of security incidents \cite{b6}. Malicious actors aim to use social engineering to exploit users into giving up valuable and confidential information \cite{b8}. We present results from a study of susceptibility of undergraduate students to phishing emails. In \textit{phishing}, a fraudulent entity tries to gain user information, possibly poising as an authority.

This observational study is the first to examine age, gender, college affiliation, academic year progression, time spent on a computer, cyber club/cyber scholarship program affiliation, cyber training, and phishing awareness demographics in one study. Our motivation lies in understanding dependent variables in a student population for future training tailored to individual students. We hope our results will help businesses and colleges improve their cybersecurity practices. 

As summarized in the tables and figures, our contributions are the 
correlations among demographics and phishing susceptibility
from our observational study in which we sent phishing emails to 1,350 UMBC students. 
For more details, see Diaz \cite{b15}.

\section{Previous Work}

There have been few phishing and general cybersecurity related surveys conducted on college students in the past, focusing on the correlation between susceptibility and one or few demographics. 

Farooq, et al. \cite{b17} studied 1280 participants in six different colleges throughout India, Malaysia, Nepal, Pakistan, and Thailand. They documented Internet use and its correlation to the student user susceptibility level. A year prior, Farooq, et. al. \cite{b16} also surveyed 614 university students from eight different majors to calculate their information security awareness score (ISA). They concluded that gender provides an insight on how a student learns cybersecurity skills. Men tend to gain security knowledge through self-taught means, while women tend to prefer formal training and interacting in their social circles \cite{b17}.  

In Tamil Nadu, India, Senthilkumar and Easwaramoorthy~\cite{b19} surveyed student responses to cyber themes, such as ``virus[es], phishing, fake advertisement, popup windows and other attacks in the internet" \cite{b19}. 
In this study, only 10 of the 379 students stated that they would report any malicious activity to their cyber crime office. 
Similarly, Kim~\cite{b18} surveyed a group of business undergraduate students on their knowledge of cyber related topics . 
While the students were somewhat knowledgeable on most topics covered in NIST Standard 800-50, Kim~\cite{b18} suggested training programs for all students within the college to increase student awareness. 
Duggan~\cite{b20} conducted a comparable survey in Japan, where he surveyed a group of Japanese college students about their cybersecurity and privacy-risk knowledge based on terminology. 

Dodge Jr., et al. \cite{b3} conducted an unannounced phishing test on students at the United States Military Academy to evaluate their cyber training programs. They concluded that the more educated a student was in academic year, the less likely they were to fall for the phishing scam. Similarly, Aloul \cite{b1} presented a project in which a fake website portal recorded the number of students who navigate to this phishing trap. They recorded 9\% of the 11,000 students falling for the fraudulent portal. 

Sheng, et al. \cite{b11} studied if age, sex, and education level influenced phishing susceptibility. They determined that higher education level, age, and being male lead to less susceptibility. Sun, et al. \cite{b12} investigated links between gender and behavior. In contrast, the research team did not find a significant difference in gender. In these two studies, the users knew that they were being tested on their ability to detect phishing attacks. 

In our study we include a more expansive list of demographics than those explored in previous studies.  We also focus on phishing susceptibility rather than on general cybersecurity topics, and we do not inform the participants beforehand of the phishing experiments. 

\section{Experimental Methodology}
\label{sec:method}

We deploy three phishing experiments on randomly-selected students at UMBC. To simulate errors found commonly in phishing attempts, these phishing emails contain errors that provide clues of their illegitimacy. Subsequently, we sent a debriefing statement to all selected students, as required by our UMBC Institutional Review Board (IRB) approval. We also sent a survey to gather more demographic data to those students who had opened a phishing email. 

\subsection{Subject Population}

Our study takes the 11,234 undergraduate students currently enrolled at UMBC as the target pool \cite{b13}. UMBC is especially well known for science and technology. UMBC includes three colleges: the College of Arts, Humanities, and Social Sciences, the College of Engineering and Information Technology, and the College of Natural and Mathematical Sciences. Our study focuses on the student's primary major, regardless of any subsequent major, minor, or certificate program \cite{b13}.

We sent each phishing email to a randomly-selected set of 1,350 students. Each set comprised 450 students, with 150 students from each college.  

We decreased the number of eligible students from 11,234 to 10,920, marking students ineligible if they had an undecided major or if they were part of the interdisciplinary studies track. Interdisciplinary Studies majors have multiple majors in potentially different colleges. 

\subsection{Experiment 1: PayPal}
Experiment 1 deployed the popular \textit{Billing Problem} tactic [5]. The fraudulent entity claims to be PayPal, a popular online payment company. The email tries to entice the user to click on the email link by claiming to have received an order from them and therefore billing their PayPal account. 

There are several red flags that indicate this email is illegitimate. Atomic Empire Designs is a fake company with invalid customer service email and phone number. The Shipping Address is vague, and the zipcode is incorrect for the Baltimore region. The email timestamp is for a future time, and the total amount of money owed does not add up to the subtotal, plus tax and shipping expenses. The last line of the email stating that ``Paypal is located at ..." lists an incorrect and invalid address. Another flag is the sender's email address: any email from the PayPal business will have a ``@paypal.com" address, not ``gmail.com". The link described as Order Details is also suspicious. If one hovers over the link, it does not indicate any association with PayPal; instead, it goes through a tracking url that contains a ``thisisnotmalware" string. 

\begin{figure}[!htb]
	\centering
	\includegraphics[width=\linewidth]{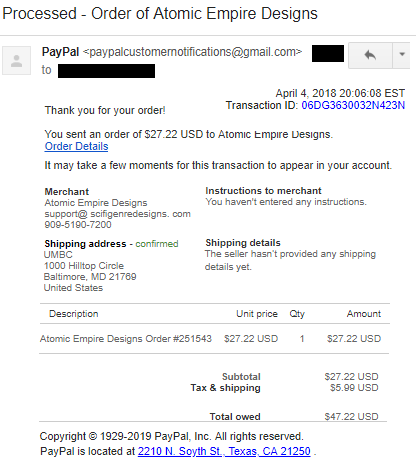}
	\caption{ Experiment 1 PayPal email claims to bill the student's PayPal account.}
\end{figure}

\begin{figure}[!htb]
	\centering
	\includegraphics[width=\linewidth]{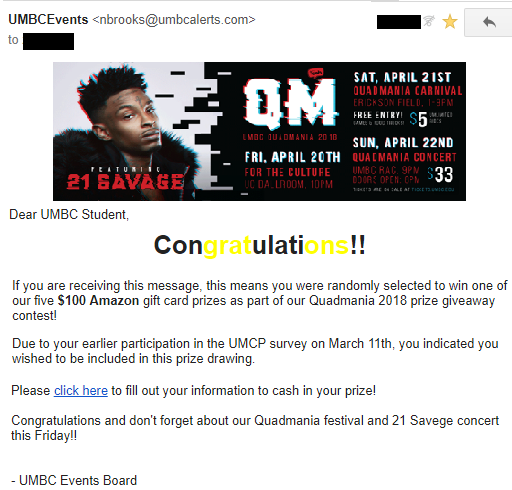}
	\caption{ Experiment 2 Quadmania email offers a free \$100 gift certificate.}
\end{figure}

\subsection{Experiment 2: Quadmania}

In this Experiment we make use of UMBC's Quadmania event, the university's major spring weekend festival, to lure the user through monetary gain \cite{b5}. The email congratulates the student on their \$100 Amazon prize and asks them to click the provided link. This email adds legitimacy by using the 2018 Quadmania banner while the signature of the email proclaims it was sent by the UMBC Events Board. This name is similar to the Student Events Board (SEB) that organizes Quadmania.  Futhermore, the email describes a UMCP survey. Not only was no such survey conducted,  UMCP refers to the University of Maryland, College Park, which is a different school. There are grammar and spelling inconsistencies, including the keynote singer 21 Savage. When hovering over the link, The user can see the link redirects them to cnn.com after going through a tracking software. The email is sent from a ``@umbcalerts.com" address, instead of a ``umbc.edu" address.

\subsection{Experiment 3: DoIT}

This email is a variation of the \textit{Expiration Date} tactic, mimicking UMBC's Division of Information Technology (DoIT). It claims that the user must verify their credentials to keep their UMBC account, referencing the Quadmania phish to add validity. The email threatens that the user must click and verify their identity within 48 hours.

There are several spelling and grammar errors, which are uncommon for official UMBC communications. The authority names itself ``Department of Institutional Technology," and later signs off with ``UNCP DoIT". There is no Department of Institutional Technology nor UNCP entity at UMBC. The odd quote at the end of the email is out of character and unconventional for a university's IT department. The email address and link of this email are suspicious as well. The user can hover over the link and see that it goes to the Google homepage after going through tracking software. The email address has a ``@umbcdoit.com" email address instead of a ``@umbc.edu" one.

\begin{figure}[!htb]
	\centering
	\includegraphics[width=\linewidth]{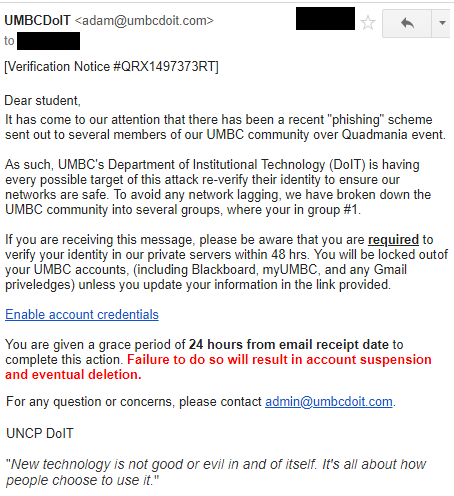}
	\caption{ Experiment 3 DoIT email threatens to suspend the student's computer account.}
\end{figure}

\subsection{Debriefing Statement and Demographic Survey}

Part of our IRB protocol requires us to send a debriefing email that informs all 1,350 selected students of the study. It also assures that we anonymized all data, kept all data confidential, and could not identify any individual. 

We invited students who were part of the 1,350 target group and opened a phishing email from Experiments 1-3 to participate in a survey. After asking for consent and ensuring the survey respondents were at least 18 years of age, we asked questions on their academic year, major affiliation, gender, age, past cybersecurity training, participation in cyber clubs or cyber scholarship programs, phishing awareness, and time spent per day on the computer. We gave a brief definition of phishing and quick tips on how to identify phishing emails. We directed the users to the official UMBC phishing and spam FAQ page for more information.

\subsection{Data Collection}
To track the data, we used the free application MailTracker by Hunter and the EmailTracker by cloudHQ  \cite{b2}\cite{b7}. Each of these programs tracked if an email recipient opened an email and whether they clicked any links. Both verify and confirm each other's recorded data.
 
\subsection{Statistical Methods}
We applied Fisher's Exact test and Pearson's Chi-Square for significance testing, and Cramer's V to test strength of that significance, with $\alpha = 0.05$ \cite{b14}. We used Fisher's Exact test in lieu of the Chi-Square test when an expected value is less than~5. We defined the null hypothesis as there is no dependency between the demographic factor and student click rate. We used IBM's SPSS to create contingency tables and calculate these statistics.  

\section{Results}
\begin{figure*} [!h]
	
	\centering
	\includegraphics[width=\linewidth]{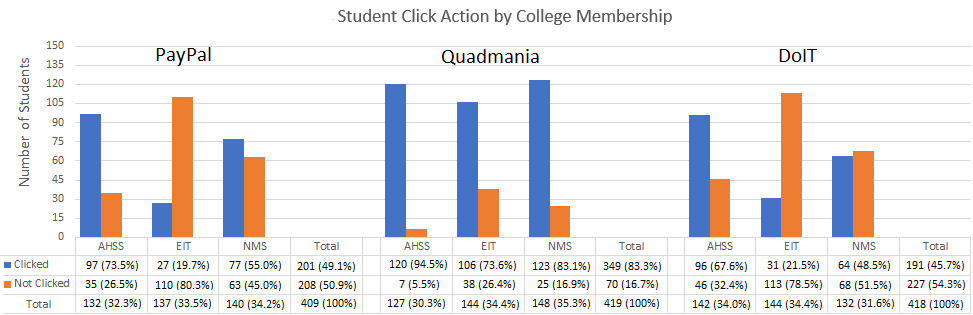}
	\label{fig:EntireColleges}
	\caption[ Overall College Data]{Number of clicks on phishing emails by students in the College of Arts, Humanities, and Social Sciences (AHSS), the College of Engineering and Information Technology (EIT), and the College of Natural and Mathematical Sciences (NMS).}
\end{figure*}

Of the 1,350 students randomly selected for this study, 1,246 (92\%) opened a phishing email in at least one of the three experiments. We sent the debriefing statement to all 1,350 students, and the demographic survey only to those 1,246 students who opened a phishing email. 
Except for college affiliation, we analyzed demographics only from survey respondent data.

\subsection{Experiment 1 Results}

Of the 450 students receiving the PayPal phishing email, 409 (91\%) opened the email. Of those 409 students, a majority of the Arts, Humanities, and Social Sciences majors clicked the link.

We sent emails to 150 students within each college and analyzed the actions of those who opened the email. 74\% of students in Arts, Humanities, and Social Sciences majors had clicked the link, with 20\%  in Engineering and Information Technology and 55\% in Natural and Mathematical Sciences. 

\subsection{Experiment 2 Results}

We sent the Quadmania phishing email to 450 students, of which 419 (93\%) opened the email. 349 students (83.3\%) clicked the link within the email. Almost all of the Arts, Humanities, and Social Sciences majors clicked the link (95\%), often within minutes of receiving the email. 74\% of students in the College of Engineering and Information Technology clicked the link, while 83\% in the College of Natural and Mathematical Sciences clicked.

\subsection{Experiment 3 Results}

93\% of students opened the third email. 68\% of students in the Arts, Humanities, and 49\% Social Sciences and Natural and Mathematical Sciences were fooled into clicking the link. In contrast, only 31 students (22\%) in Engineering and Information Technology majors clicked.  

\subsection{Survey Results}

Of the 1,246 students who had the option to complete the survey, 482 students (39\%) responded within a seven-day period. For each cohort, at least 100 subjects completed the survey. Figure 4 shows the click action by college membership for each experiment.

\begin{figure*} [!h]
	\centering
	\includegraphics[width=\linewidth]{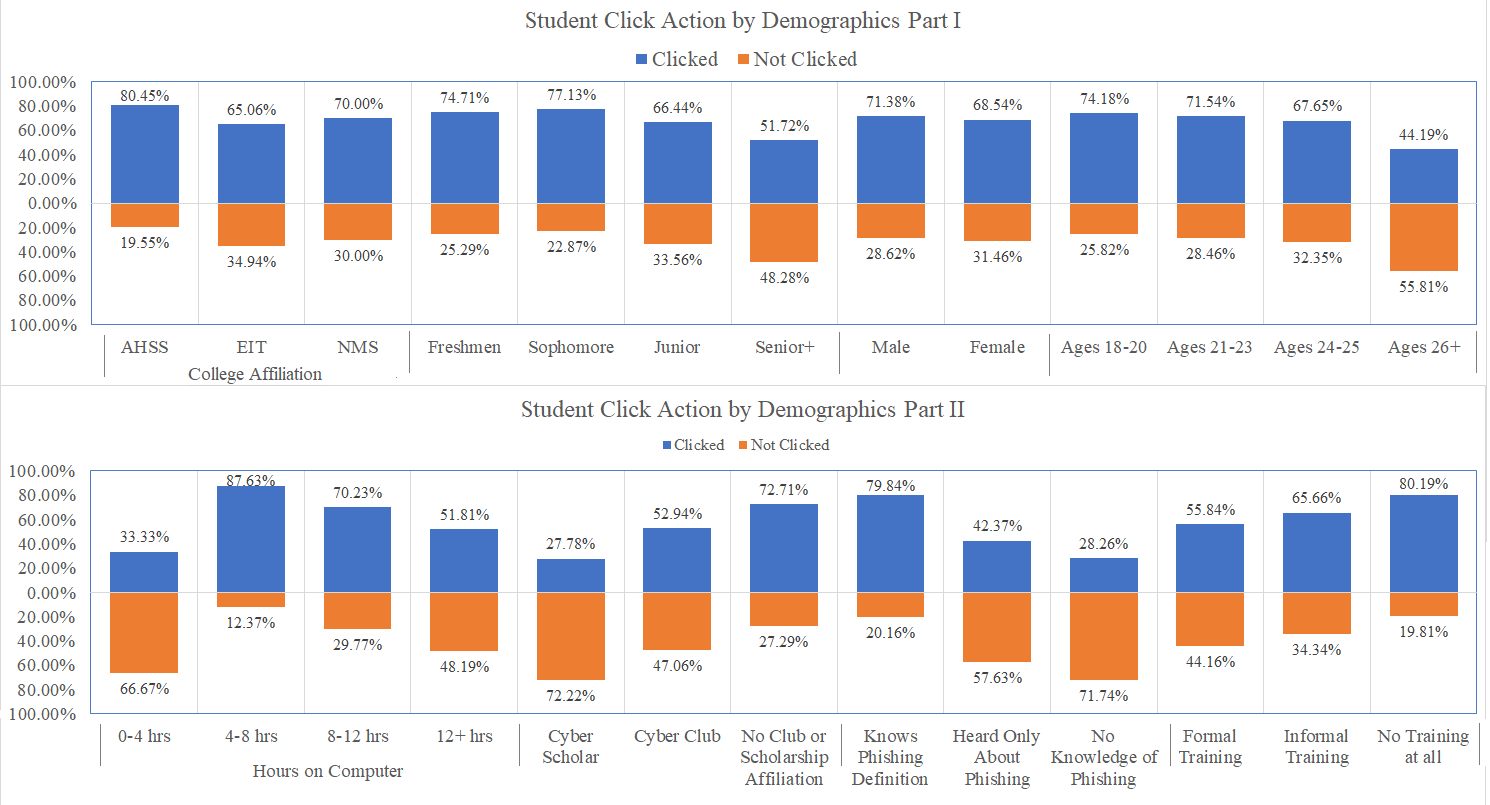}
	\label{fig:stackedbar}
	\caption[Significant Demographic Data]{ Click action by demographic factors for students who opened email and returned the demographic survey form.}
\end{figure*}

\section{Analysis}

We analyze all experiments and survey results and find significant correlations in all tested demographics except gender. 

Table~1 lists the number and percentages of students who clicked on phishing emails, 
with the results listed separately for students who
were sent emails, opened emails, and answered the demographic survey.

\begin{table}[!h]
	\caption[Survey Experiment Breakdown]{Summary of experimental results. Number of students who clicked on phishing emails, among students who were sent emails, opened the emails, and answered the survey.}
	\centering
	\includegraphics[width=\linewidth]{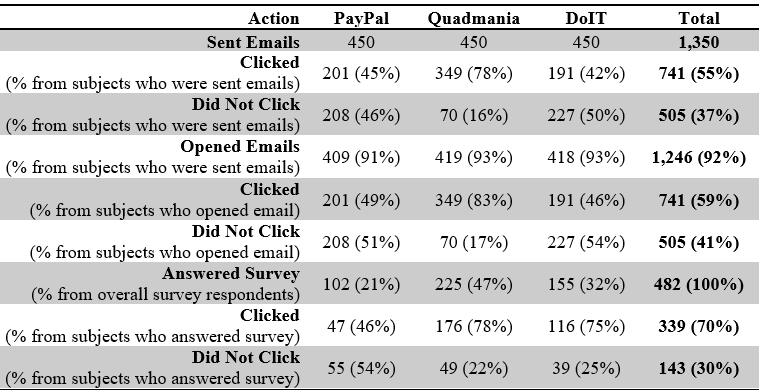}
	\label{fig:SurveyTable}
	
\end{table}

Approximately 59\% of the subjects who opened the email clicked on the phishing link, with
some fluctuations among the three experiments. 
By contrast, approximately 70\% of the survey respondents clicked.

\subsection{Experiments}
\begin{table}[!htb]
	\caption[Experiment Statistical Tests]{Significance of three statistical tests at separating students who click on emails, computed separately for each phishing email, at confidence level $\alpha = 0.05$, with given degrees of freedom (df).}
	\centering
	\includegraphics[width=\linewidth]{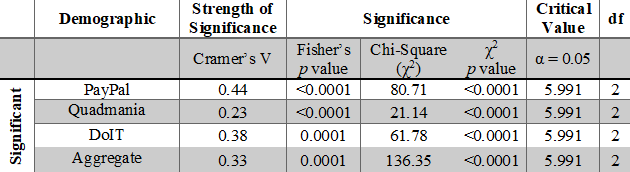}
	
\end{table}

We found a correlation between college affiliation and user click action. For all three experiments, the Chi-Square value exceeded $5.991$. The aggregate data also had a Chi-Square value exceeding the critical value, rejecting the null hypothesis. We define the null hypothesis as there being no correlation between user susceptibility and a demographic. A low-to-medium strength of association is also present. 

\subsection{Comparative Analysis}

We show that phishing awareness, hours spent on the computer, cyber training, cyber club or cyber scholarship affiliation, age, academic year, and college affiliation are significant variables to student susceptibility. 

\begin{table}[!htb]
	\caption[Experiment 3 Demographic Significance Tests]{Significance of three statistical tests at separating students who clicked on a phishing email, by demographic factors, at confidence level $\alpha = 0.05$, with given degrees of freedom (df) .}
	\centering
	\includegraphics[width=\linewidth]{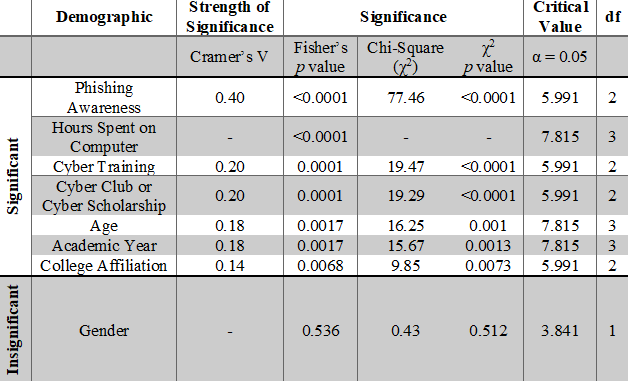}

\end{table}

The aggregated college affiliation demographic indicates that STEM majors--with Engineering and IT majors in particular--had lower click rates (EIT 65\%, NMS 70\%) compared to non-STEM majors (AHSS 80\%). Increasing academic year progression saw a decrease in student click rate. We observed that increased time on the computer and cyber training correlated with lower click rates. Students in a cyber club or cyber scholarship program also clicked the phishing link less often than did students with no such affiliation. Within the cyber club and cyber scholarship group, students who were affiliated with a cyber scholarship program had lower click rates compared to the cyber club students. 

Contrary to our expectations, in Experiments 1--3, students who were unaware of phishing attacks performed better (28\% clicked) than did students who were aware (42\% clicked) or who understood what phishing attacks are (80\% clicked).

We found no significant correlation between gender and susceptibility, with the Chi-Square calculation less than the critical value.

\section{Discussion}

We describe the campus response to our phishing emails,
discuss an unexpected finding, 
comment on the nature of the phishing emails, 
identify study limitations, 
and present open problems.

\subsection{Campus Response}

Although the PayPal email received little attention, the Quadmania phish (purportedly from SEB)
created notable confusion.
SEB, DoIT, and campus police issued alerts. 
A few hours after we sent the emails, SEB posted warnings to the student body of a phishing scheme,
informing users that they did not send the Quadmania email, spreading word on 
the \textit{myUMBC} dashboard and social media. 

SEB's quick and efficient communication reached several students within the Experiment~2 cohort. 
Despite these warnings, the vast majority of students had already fallen for the Quadmania scheme. 
Many students who were deceived by the phish reported their experiences to DoIT or SEB, 
prompting quick responses by SEB and DoIT to us and the student body.  
While we had notified DoIT in advance, not all of their staff knew about our experiment, and
in hindsight, we probably should have also informed SEB in advance.

\subsection{An Unexpected Finding}

As expected, we observed lower user susceptibility with
college affiliation, academic year, age, cyber club and cyber scholarship affiliation, 
amount of time spent on the computer, and cyber training. 
Contrary to our expectations, we observed greater user susceptibility with greater phishing knowledge and awareness. 

We have no convincing explanation for this finding, and we do not know if it is reproducible.
Nevertheless, we consider two speculations.
First, it is possible that the act of falling for the phishing scheme might have increased the user's 
awareness about phishing.  In hindsight, it might have been wiser to have asked in the post-event survey
what was the level of phishing awareness the user had when they opened the phishing email.
Second, it is conceivable that users who fell for the phish might be more likely to overestimate their 
knowledge, including about phishing.

\subsection{Limitations}

Limitations of the study include student awareness of the experiment and veracity of survey responses.
Especially given the commotion created by the Quadmania phish, it is possible that there was greater
awareness among subjects about the possibility of phishing attacks in the third experiment than in the first two.
We made no attempt to measure how accurately and honestly subjects filled out their demographic surveys.

We can analyze the correlation between demographics and suseptibility only for subjects who answered the demographic survey. 
It is likely that survey respondents are a somewhat biased sample of the undergraduate population---for example, survey respondents
might be less cautious and more likely to act on opportunities.  
This bias possibly explains why survey respondents clicked on the phishing link at a higher rate than did students who merely opened the phishing email.

\subsection{Nature of Phishing Emails}

As explained in Section~\ref{sec:method}, we intentionally inserted many clues into each phishing email of their illegitimacy (e.g., spelling errors), and initially, we did not inform the subjects about the experiments.  Our rationale was to simulate commonly occurring phishing attacks, which often contain such clues.  We do not know how much, if it all, such clues affected user behavior.  Similarly, we do not know how much, if at all, lack of awareness of the experiment affected user behavior.  Given the high click rates, we speculate that, for many users, such clues were not a decisive factor.  Similarly, given that study awareness appears to be a more subtle issue, and that many users are generally aware about the possibility of phishing attacks, we speculate that lack of awareness of study did not make a significant difference.

Alarmingly, given the high click rates for our phishing emails with many clues, we believe that most users would be even more highly susceptible to more sophisticated attacks.  In a more sophisticated attack, the adversary might surveil the target and construct a compelling customized spear-phishing email free of any obvious clues.

\subsection{Open Problems}

It would be interesting to understand our unexpected finding that students who reported greater 
phishing knowledge were more susceptible.  Additional studies could explore this question and determine
if our findings are reproducible. It would be useful to understand how clues and study awareness affect user behavior.  
It would be interesting to include faculty and staff in a study and
to analyze user behaviors over several semesters. More difficult open problems are to explore causal factors
in user behaviors and to devise effective ways to combat the threat of phishing attacks, including
better user education, email filtering, and system design.

\section{Conclusion}

Our study finds an association between several demographic factors and a student's susceptibility to phishing attack. We observed lower susceptibility for college affiliation, academic year progression, cyber training, involvement in cyber clubs or cyber scholarship programs, amount of time spent on the computer, and age demographics. Surprisingly, 
despite a lower susceptibility for cyber education or IT expertise,
we observed greater susceptibility for phishing awareness. 
We found no significant correlation for gender. 

Phishing attacks are a dangerous form of social engineering that target users every day. 
Our study shows that user susceptibility to phishing remains a prevalent problem, even among technology-savvy students: 
approximately 59\% of the subjects who openened the phishing email clicked on the phishing link,
and approximately 70\% of those subjects who also answered the demographic survey clicked.
Our observational study uncovers relationships between
demographic factors and susceptibility to phishing.  We hope that these findings will be helpful in designing
more secure systems and developing more effective cybersecurity training for users.

\section*{Acknowledgments}
The authors thank Professors Bimal Sinha and Nagaraj Neerchal for their counsel on statistical tests and models. We would also like to thank Jack Seuss, Andy Johnston, Mark Cather, and the DoIT staff for their support and help throughout the project. 

Sherman was supported in part by the National Science Foundation under SFS grant 1241576 and by the U.S. Department of Defense under CAE grant H98230-17-1-0349. Joshi was supported by an award from IBM.

\vspace{12pt}

Submitted on November 12, 2018, to \textit{Cryptologia}.

\end{document}